\documentclass[prb,aps,superbib,twocolumn,floatfix,superscriptaddress]{revtex4}
\usepackage{times,amsfonts,amsmath,amssymb,graphicx}
\begin{document}
\title{Microscopic Current Dynamics in Nanoscale Junctions}
\author{Na Sai}
\affiliation{Department of Physics, University of California, San
Diego, La Jolla, CA 92093-0319}
\affiliation{Department of Physics, The University of Texas, Austin, TX 78712}

\author{Neil Bushong}
\affiliation{Department of Physics, University of California, San
Diego, La Jolla, CA 92093-0319}

\author{Ryan Hatcher}
\affiliation{Department of Physics, Vanderbilt University, Nashville,
TN 37235}

\author{Massimiliano Di Ventra}
\affiliation{Department of Physics, University of California, San
Diego, La Jolla, CA 92093-0319}

\date{January 07, 2007}
%\date{\today}
\begin{abstract}

So far transport properties of nanoscale contacts have been mostly studied
within the static scattering approach. The electron dynamics and the
transient behavior of current flow, however, remain poorly understood. We
present a numerical study of microscopic current flow dynamics in nanoscale
quantum point contacts.  We employ an approach that combines a microcanonical
picture of transport with time-dependent density-functional theory.  We carry
out atomic and jellium model calculations to show that the time evolution of
the current flow exhibits several noteworthy features, such as nonlaminarity
and edge flow. We attribute these features to the interaction of the electron
fluid with the ionic lattice, to the existence of pressure gradients in the
fluid, and to the transient dynamical formation of surface charges at the
nanocontact-electrode interfaces. Our results suggest that quantum transport
systems exhibit hydrodynamical characteristics which resemble those of a
classical liquid.

\end{abstract}

\maketitle
\section{Introduction}
\label{sec:intro}
Recent experimental progress has enabled imaging of coherent current flow
dynamics in quantum point contacts formed in semiconductor
heterostructures.\cite {Briner96p5283, Eriksson96p671, Feenstr98p699,
Topinka00} These advances in experimental techniques open the possibility
that current flow through atomic or molecular junctions will be eventually
imaged and controlled. Understanding the microscopic electronic flow patterns
can aid the design of novel electronic devices. However, very few theoretical
studies of current dynamics in nanoscale systems are currently available.
Indeed, among the recent theoretical studies of transport in nanoscale
systems, much emphasis has been placed on the steady-state conduction
properties,\cite
{Emberly98p5810911,DiVentra00p979,Xue01p4292,Taylor01p245407,Damle01p201403,Brandbyge02p165401,Heurich02p6803, Palacios02p035322,Pecchia04p1497,Rocha06p085414,Prociuk06p4717}
whereas the transient behavior of the current remains an unexplored area.

Electronic transport in nanoscale junctions is usually formulated within the
stationary scattering picture, such as the one due to
Landauer,\cite{Landauer57} in which the conduction is treated as a collection
of scattering events. This stationary approach, widely used to study
transport in mesoscopic and nanoscale systems, has led to considerable
success in understanding current-related effects other than the conductance,
including, e.g., noise, local heating, and current-induced
forces.\cite{curr-effects} It has helped our understanding of the microscopic
current distribution as well.\cite{Lang87p8173,Todorov99p1577} Nevertheless,
the stationary approach assumes that the system is already in a steady state,
leaving the questions of how a steady current establishes itself and what
other phenomena are related to the dynamical formation of steady states in a
nanojunction unsolved.

The dynamical nature of the current flow is better addressed in a
time-dependent approach than in the stationary one. Time-dependent or AC
transport approaches have been previously introduced in mesoscopic conducting
systems.\cite{Wingreen93p8487,Grifoni98p229,Buttiker00p519} Recently, an
increasing amount of effort has been directed toward developing ab-initio
time- dependent approaches for nanoscale systems.\cite{DiVentra04p8025,
Baer04p3387,
Stefanucci04p195318,Horsfield04p8251,Kurth05p035308,Bushong05p2569,
Burke05p146803, Zhu05p075317, Pedersen05p195330, Qian06p035408,
Sanchez06p214708} One such method was developed to study the AC conductance
using the time-dependent density functional theory\cite{Runge84p997}(TDDFT)
combined with absorbing boundary conditions.\cite{Baer04p3387} This method,
however, is affected by the arbitrariness of how and where the absorbing
potentials are added, while the effect of the absorbing potential on the
conduction in nanojunctions is unclear. Other methods have been developed
based on the Landauer scattering formalism of transport that employ open
boundary conditions.\cite{Stefanucci04p195318, Kurth05p035308,
Burke05p146803, Zhu05p075317}

More recently, a microcanonical formalism that treats electronic transport as
a discharge across a nanocontact connecting two large but finite charged
electrodes was introduced.\cite{DiVentra04p8025} The formulation has been combined with TDDFT to study the dynamical formation of quasi-steady currents, local
chemical potentials,\cite{Bushong05p2569} and electron-ion
interactions.\cite{Horsfield04p8251} This formalism would yield the exact
total current flowing from one electrode to the other if the exact functional
were known, regardless of whether the system achieves a steady state. In
practice, in ab-initio transport calculations, one only uses approximate
forms of the functional such as the adiabatic local density approximation
(ALDA). It has been shown recently that the electronic correlation effect
beyond the ALDA gives rise to additional resistance in molecular
junctions.\cite{Sai05p186810,Koentopp06p12l1403,Bokes0604317} The spurious
self-interaction implicit in the ALDA further complicates calculations of the
conduction properties.\cite{Toher05p146402,Ke0609637,Muralidharan06p155410}
The sensitivity of various transport properties to the suggested corrections
remains only partially understood.

The dynamical establishment of a quasi-steady current has been investigated
by a number of authors.\cite{Bushong05p2569, Kurth05p035308,
Horsfield04p8251,Stefanucci0608401} It has been demonstrated that a quasi-steady current can
establish itself across a junction on a femtosecond time scale without the
presence of inelastic scattering.\cite{Bushong05p2569} This is due to the
geometrical ``squeezing'' experienced by the electrons crossing the
nanojunction.\cite{DiVentra04p8025} The conductance calculated from the
microcanonical formula was shown to be in good agreement with that obtained
from the static DFT approach in prototypical atomic
junctions\cite{Bushong05p2569, Horsfield04p8251} as well as in molecular
junctions.\cite{Cheng06p155112} Nevertheless, a study of the microscopic
behavior of the electron flow, and in particular of the current flow
morphology in nanojunctions, is still lacking.

In this paper, we carry out real-time numerical simulations of current flow
in metallic nanojunctions using the microcanonical formalism, where we employ
TDDFT within the ALDA. We restrict the forthcoming discussion to the
dynamical behavior of electron/hole charges in the nanojunctions under the
linear response regime, i.e., in which the bias is small and the
current-voltage characteristics are linear. The paper is organized as
follows. In Sec.\ref{sec:methods} we discuss model transport systems and
numerical methods. In Sec.\ref{sec:results}, we present and discuss
simulations of current dynamics in jellium and atomic junctions. We also
analyze the effects of hydrodynamic pressure and electrode surface charges on
the dynamics of the flow. In Sec.\ref{sec:conclusions}, we summarize the main
conclusions of our work.

\begin{figure}
\includegraphics[width=2.2in]{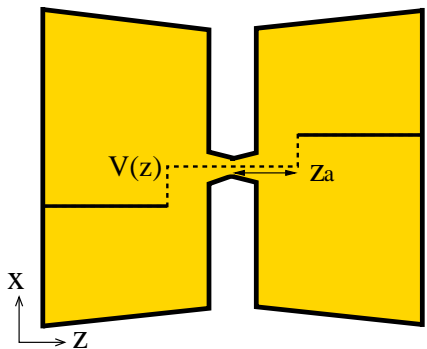}
\caption{(Color online) Sketch of the nanojunction geometry that is  
studied in the
paper. At $t<0$, a bias in the form of $V(z) = V_0[{\rm H}(z-z_a) -
{\rm H}(-z-z_a)]$ is applied to the junction (the central
constriction is at $z=0$) such that the regions $|z| > z_a$ bear a
potential offset from the central constriction, where $H(z)$ is the
Heaviside step function.}
\label{fig:schematic}
\end{figure}
\section{Model and Methods}
\label{sec:methods}
The nanoscale junction geometry studied in this paper is illustrated in
Fig.~\ref{fig:schematic}. A narrow constriction separates two large but
finite electrodes. We begin the simulations by applying a step function-like
electric bias across the junction such that the two electrodes bear equal and
opposite potentials offset relative to the potential at the center of the
junction. The distance from the bias discontinuity to the center of the
junction is $z_a$ (see Fig.\ref{fig:schematic}).  This bias induces a charge
imbalance between the two sides of the system. At $t=0$, we remove the bias
and a discharge through the nanojunction ensues. The Kohn-Sham initial state
therefore corresponds to the ground state of the Hamiltonian in the presence
of the bias. Here, we are interested in the transient behavior during the
phase in which the current is in the process of establishing a quasi-steady
state and immediately thereafter, i.e., long before the electrons that have
passed through the constriction have had chance to reach the far boundary of
the electrodes.

To separate the effects of the electrons and of the atomic lattice, we have
carried out calculations using a jellium model and an atomic model.  In the
jellium calculations, the electrodes are represented by two large jellium
slabs 2.8 \AA\ thick.  The contact is a rectangular jellium block 2.8 \AA\
wide and as thick as the electrodes. The distance between the jellium
electrodes is 9.8 \AA. In the atomic calculations, the junction is
represented by two planar arrays of gold atoms sandwiching a single gold
atom. We employ TDDFT and solve the effective single-particle Schr \"odinger
equation
\begin{equation}
i\hbar {\dot \psi}({\vec r},t) =\left[-\frac{\hbar^2\nabla^2}{2m}
+ V_{\rm eff}( {\vec r},t)\right]\psi({\vec r},t) ,
\end{equation}
where the effective potential is given by
\begin{equation}
V_{\rm eff}({\vec r},t) = V_{\rm ext}({\vec r},t) + \int\frac{\rho 
({\vec r},t)}{{\vec r}-{\vec r}'}d{\vec r}' + V_{\rm xc}({\vec r},t).
\end{equation}

The term $V_{\rm xc}({\vec r},t)$ is the exchange-correlation potential
calculated within the adiabatic local- density approximation. The external
(ionic) potential is modeled using pseudopotentials for the atomic
calculations,\cite{socorro} while in the jellium model it is a local operator
related to the uniform positive background jellium density $\rho_0$ via
$V_{\rm ext}({\vec r}) = \int \frac{\rho_{0}}{{\vec r}-{\vec r}'}d {\vec
r}'$, where $\rho_{0}$ equals to $(\frac{4\pi r_{\rm s}^3 }{3})^{-1}$ inside
the jellium and zero outside, and $r_s$ is the Wigner-Seitz radius. In the
jellium model, we choose $r_s=3a_B$ ($a_B$ = Bohr radius) which gives a good
representation of bulk gold (see also discussion below). A ``free- space''
boundary condition is implemented such that the long-range potential is
constructed only from the densities in the supercell;\cite{hockney70} that
is, the system is not periodic. Additional numerical details can be found in
Appendix~\ref{num}. The single-particle time-dependent current density is
calculated via
\begin{equation} {\vec j}({\vec r},t) = \sum_n
\frac{\hbar}{2mi}\left[\psi_n^*({\vec r},t){\vec \nabla}\psi_n({\vec  
r},t) -
\nabla\psi_n^*({\vec r},t)\psi_n({\vec r},t)\right] ,
\end{equation}
where $\psi_n$ denotes individual Kohn-Sham single-particle states. Note that
in TDDFT the current density is not necessarily exact even if one calculates
it with the exact functional (via the continuity equation, only the divergence of the current density would be exact). One would need to resort to Time-Dependent
Current Density Functional Theory (TDCDFT)~\cite{Vignale96p2037} to obtain the exact
current density (with the exact functional). Nonetheless, due to the small
viscosity of the electron liquid, we have found that the current density one
obtains by using TDDFT within the ALDA, and the one obtained by TDCDFT within
the Vignale-Kohn functional~\cite{Vignale96p2037} are qualitatively similar.~\cite{BGD}

Even if the contact between the electrodes were removed, the current between
the two electrodes would not completely vanish because of quantum tunneling.
This bare tunneling current can conveniently be used to compare the jellium
and the atomic calculations. The jellium edges are placed at half the
interplanar spacing of the lattice.\cite{Lang03} This way, the jellium model
and the pseudopotential calculations both yield tunneling current densities
of $\sim$ 0.05 $\mu$A/\AA$^{2}$ at a bias of 0.2V. The agreement indicates
that the jellium model is a good representation of two large metal
electrodes. This is consistent with the results of previous
density-functional calculations.\cite{DiVentra02}

\begin{figure*}
\includegraphics[width=7.0in]{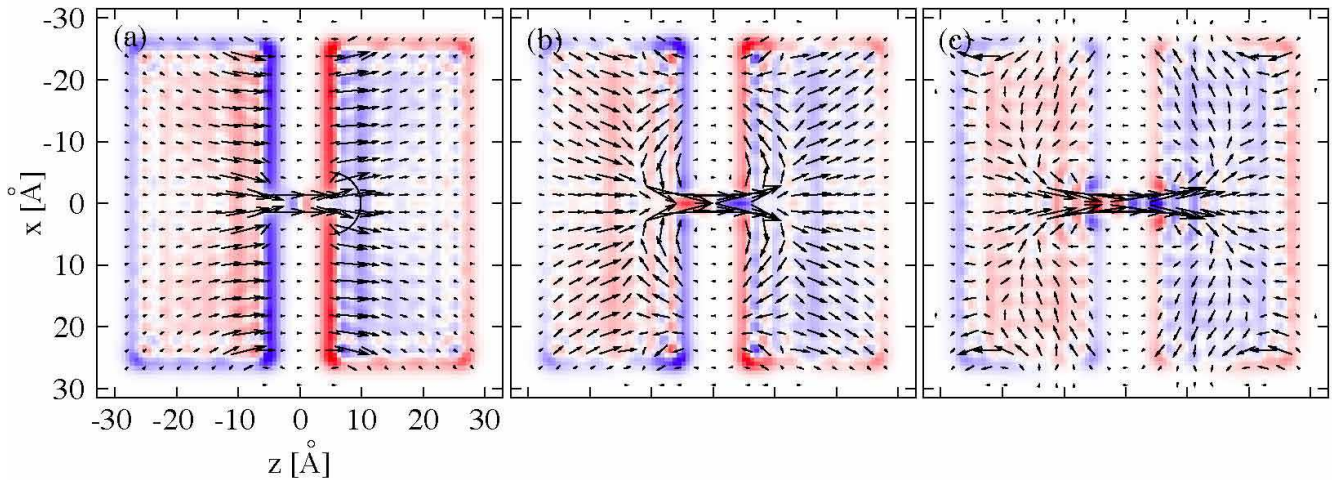}
\caption{(Color online) Current flux for a series of times in a nanoscale
quantum point contact system in the jellium model. The applied bias at $t<0$
is $ \Delta V= 0.2$V. The field lines in each panel depict the direction and
amplitude of the current density vectors, while the colors give extra
electron (red) or hole (blue) density. (a) $t=0.4$ fs ; (b) $t=0.8$ fs ; (c)
$t=1.6$ fs. In (a), the semicircle marks the contour along which the radial
component of the current density is calculated (see text and
Fig.~\ref{fig:radial_jel}).}
\label{fig:cur_jel}
\end{figure*}
\section{Results and Discussion}
\label{sec:results}
\subsection{Flow dynamics through jellium model junctions}
\label{sec:jell}
In a nanojunction such as an atomic point contact, the dimensions of the
leads are usually much larger than those of the central constriction. In
addition, not far from the contact, we expect the electron momentum to
converge to the value characteristic of the bulk leads.  Therefore, the
momentum of an electron coming from the leads and entering the contact has to
change considerably. This gives rise to resistance, and for a truly nanoscale
junction, this momentum mismatch is mainly responsible for the establishment
of quasi-steady states.\cite{DiVentra04p8025,Bushong05p2569,explain} Using
the above dynamical approach we can now study how this translates into
microscopic current flow through the nanocontact and into the leads by
calculating the current density at different times.

To begin, we follow the method described in the preceding section to impose a
charge imbalance in the jellium model system. A discussion of the effect of
the lattice on the flow dynamics will be presented in the following section.
The initial bias is chosen such that the discontinuity happens at the edge of
the jellium slab near the central constriction (i.e., $z_a\cong5$ \AA). The
flow pattern is independent of the location of the discontinuity once the
current starts to flow through the center of the junction. In Fig.~\ref
{fig:cur_jel}, we plot three snapshots of the current density to illustrate
the evolution of the flow. Due to the bias offset near the jellium edges, a
dipolar layer forms on each of the two contact-electrode interfaces.  As a
result, the initial current flow is uniform on both sides as shown in
Fig.~\ref{fig:cur_jel}(a). Very little current flows in the nanojunction at
this point, however the current steadily rises. In Fig.~\ref
{fig:cur_jel}(b), the current density becomes convergent toward the center of
the nanojunction. Interestingly, as the excess charge from the left electrode
reaches the contact, there is a period of adjustment during which the
dominant flow is in the lateral direction, i.e., parallel to the facing
surfaces of the electrodes.

\begin{figure}
\includegraphics[angle=270,width=3.5in]{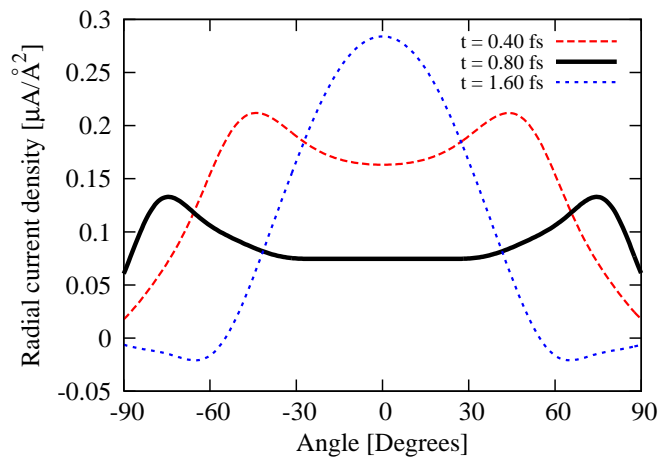}
\caption{(Color online) Time series of the radial amplitude of the current
densities along a semicircle of radius 3\AA\ centered on the nanojunction, as
a function of angle along the contour.}
\label{fig:radial_jel}
\end{figure}
\begin{figure*}[htb]
\includegraphics[angle=270,width=5.8in]{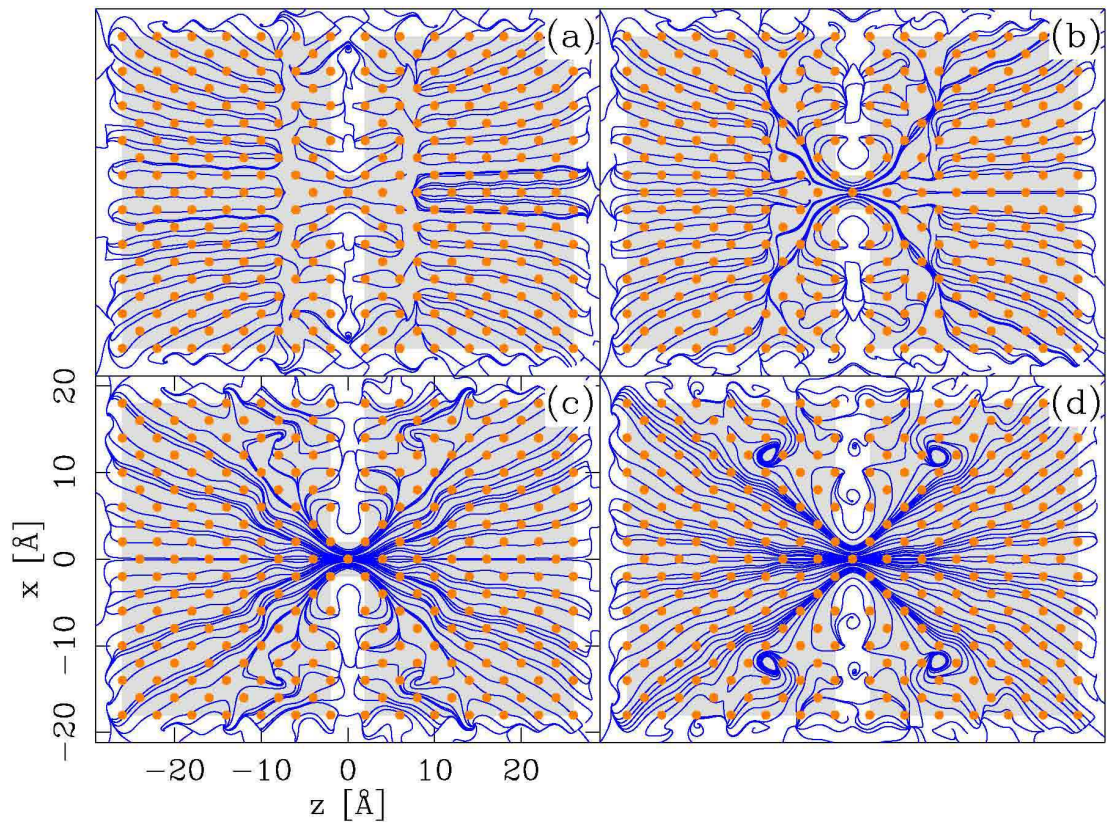}
\caption{(Color online) Time sequence of electron current streamlines  
in the atomic junction described in Sec.\ref{sec:methods}. Panels (a) - (d)
correspond to $t = $ 0.2, 0.25, 0.3, and 0.35 fs, respectively. The dots
denote atomic sites that corresponds to the (001) facets of the gold FCC
lattice. The applied bias at $t<0$ is $\Delta V = 0.2$V.}
\label{fig:cur_at}
\end{figure*}
To quantify the evolution of the angular distribution of the electron flow,
in Fig.~\ref{fig:radial_jel} we plot a time series of the radial component of
the current density along a semicircle contour centered on the junction as a
function of the angle on the semicircle (see Fig.~\ref{fig:cur_jel} (a)). One
can see that initially a ``wave'' of excess charge approaches the
nanocontact. Then, the radial current density peaks at very large angles
($\sim\pm 75^\circ$); i.e., the current density near the contact is dominated
by the flow along the electrode edges in the lateral direction. The peaks
then gradually move towards the central axis, and the current density adjusts
to a more ``focused'' pattern as shown in Fig.~\ref{fig:cur_jel}(c).  We have
also examined a junction consisting of a jellium circular ``island'' between
the electrodes. We have also observed edge flow in this case as well. The
edge flow is not a quantum interference pattern, and cannot be compared with
the fringes observed in the 2D electron gas quantum point
contacts.\cite{Topinka00} Instead, we suggest that the flow pattern is
controlled by hydrodynamic effects and forces due to surface charges. We
analyze these in Sec.~\ref{sec:surf}.

The structure of the current density is analogous to a classical fluid
flowing across a narrow constriction. This is not surprising because an
inhomogeneous electron system can be indeed characterized by a set of
hydrodynamical relations expressed in terms of the particle density and
velocity field.\cite{Conti99p7966,Martin59p1342,Tokatly,D'Agosta05} More recently,
in particular, a hydrodynamical approach was proposed for nanoscale transport
systems,\cite{D'Agosta05} further strengthening the analogy between the
dynamics of the electron liquid and the one of a classical liquid. We note
that the present calculations do not take into account the physical viscosity
of the electron liquid.\cite{Conti99p7966,note:vis} An inviscid fluid can
therefore be used as a model for the dynamical behavior of the present
electron liquid.
\subsection{Flow dynamics through atomic junctions}
\label{sec:lattice}
The jellium model was a convenient way to probe the microscopic current
dynamics in an electron gas. To understand the influence of the lattice on
the flow, we carried out simulations that included an ionic background
modeled by pseudopotentials. The atomic calculations were carried out for 2-D
gold nanojunctions and were initialized in the same way as the jellium
calculations. The theoretical and experimental conductance are in very good
agreement for this system.\cite{Agrait03p81} We chose lattice arrangements
that correspond to the (001) and (111) facets of the gold FCC lattice (the
dots in Fig.~\ref{fig:cur_at} represent the atomic sites for the (001)
configuration).

Electric current streamlines at different times in the simulation are plotted
in Fig.~\ref{fig:cur_at}(a) - (d). The streamlines are calculated by
integrating the current density field upstream and downstream, $d {\vec r}/ds
= \pm {\vec j}({\vec r}(s))$. The morphology of the current flow in the
atomic junction and in the jellium model is remarkably
similar,\cite{note:streamline} indicating that the jellium model is a good
representation of the gold electrodes.  Nevertheless, a number of new
features appear in the atomic calculations.

Fig.~\ref{fig:cur_at}(c) shows that once a steady flow through the junction
is established, the current spreads into a wedge-shaped region inside the
electrodes. The flow morphology for each of the two different lattice
arrangements is similar except that the flow spreads over a broader
wedge-shaped region in the (111) lattice. Another common feature in the
atomic calculations is the presence of a stagnant zone at the corner of the
electrode boundary. There is little current flow into or out of this zone.
This is similar to a classical fluid where a stagnant zone can sometimes be
located at the entrance or exit of a channel.

One profound difference between the atomic and the jellium calculations is
the formation of eddies evident in the former but not in the latter.  In the
jellium calculations carried out within the linear-response bias regime, the
current flux lines are laminar. In contrast, in the atomic calculations, the
eddies appear as localized circular flow that can be observed in
Fig.~\ref{fig:cur_at}(d). The eddies develop in both electrodes and the size
of the eddies is comparable to the interatomic distance. The eddies are
reminiscent of the vortices that form in a classical fluid at higher Reynolds
number when the fluid encounters obstacles. As is well known, vortices can
also occur when velocity shear is present within a continuous fluid
(Kelvin-Helmholtz instability). We suggest that the lattice ionic obstacles
and the boundaries separating the flow zone and the stagnant zone facilitate
the formation of the observed eddies in our simulations.

The formation of current vortices has been previously reported in 2D
ballistic quantum billiards.\cite {Berggren02p016218,Zozoulenko03p085320} In
these quantum systems, a rich variety of flow patterns ranging from regular
to chaotic is possible. While we cannot draw direct analogies with these open
and mesoscopic transport systems, an unstable and turbulent flow has been
recently predicted in nanoscale systems\cite {D'Agosta05}. We have argued
that the ALDA electron liquid in our simulations corresponds approximately to
an inviscid fluid. We therefore suggest that in the presence of a lattice,
hydrodynamical instabilities, or turbulence can occur in nanotransport
systems even in the absence of a physical viscosity.
\subsection{Hydrodynamics and the formation of surface charges}
\label{sec:surf}
\begin{figure}[htb]
\includegraphics[width=3.5in ]{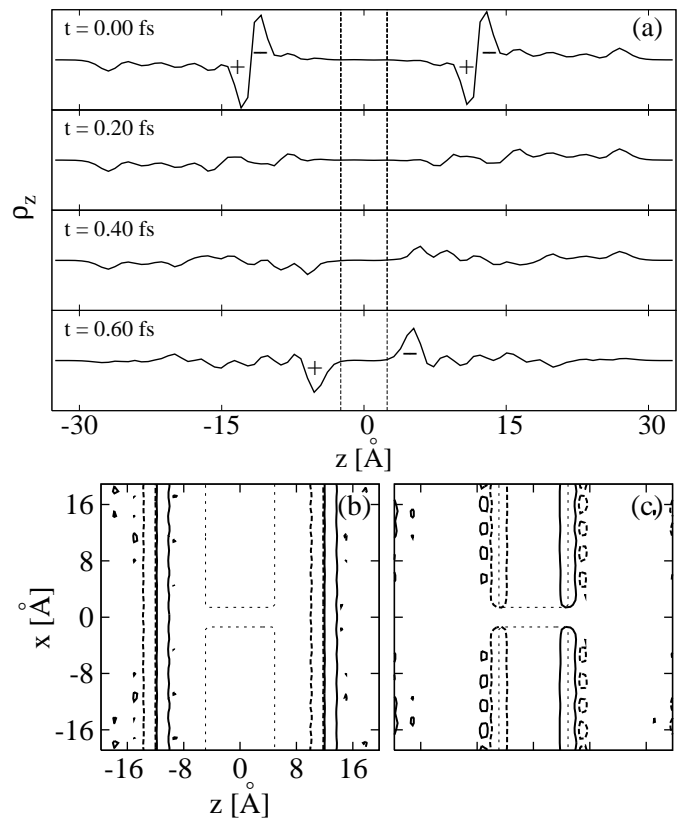}
\caption{(a) Planar averaged charge density from $t=0$ to $t=0.6$ fs for a
jellium junction. The change in the peaks indicates the dynamic process in
which excess charge builds up at the surfaces. The sign of the surface
charges indicates that electron charges accumulates on the right and hole
charges on the left.  (b) - (c) Excess charge in the vicinity of the contact
at $t=$ 0 and $t=$ 0.6 fs. Thick solid lines indicate excess electrons, while
thick dashed lines indicate excess holes.  Thin dashed lines mark the edges
of the jellium electrodes and the contact.}
\label{fig:nz_contour}
\end{figure}

To understand what drives the edge flow along the electrode surfaces, we
examine the evolution of the charge distribution near the surfaces of the junction. For this purpose,
we apply a step-function bias such that the potential discontinuity in each
electrode occurs at $z_a\cong10$\AA, which cuts across the electrode, as
illustrated in Fig.~\ref{fig:schematic}. In Fig.~\ref {fig:nz_contour}(a), we
plot a time series of the $x-y$ plane-averaged excess charge density along
the $z$ axis. At $t=0$, two symmetric dipolar layers form inside the
electrodes as a result of the bias offsets. As the current starts to flow
through the contact and gradually reaches a quasi-steady state, a global
charge redistribution becomes apparent. The dipolar layers gradually vanish
and are replaced by surface charge layers that form at the contact- electrode
interfaces. The charge contour plots Fig.~\ref{fig:nz_contour}(b-c) further
illustrate the formation of surface charges as a result of current flow.

The formation of surface charges around the central constriction is
reminiscent of the formation of residual-resistivity dipoles introduced by
Landauer.\cite{Landauer57} It has been suggested that a continuous current
flow arriving at a junction must be accompanied by self-consistently formed
charges at the electrode surfaces.\cite{DiVentra02} The effect should be
taken into account to correctly characterize the electrostatic potential and
the nonequilibrium conducting properties in a transport
calculation.\cite{Mera05p085311} In this work we provide the first numerical
demonstration of the dynamical formation of these surface charges using a
time-dependent approach.

We have already observed that, as the surfaces of the electrodes are
populated by excess charges, a lateral flow starts to develop along the
surfaces. This behavior is illustrated in Fig.~\ref{fig:radial_jel} where the
radial current flux at $t=0.8$ fs shows two pronounced peaks at very large
angles. We attempt to interpret this behavior within the framework of an
effective classic hydrodynamic model of an inviscid charged fluid. The acceleration of the fluid is then given by Euler's equation
$\partial {\vec v}/\partial t + ({\vec v}\cdot {\vec \nabla}) {\vec v} =
-{\vec \nabla} P/m_en - e{\vec \nabla} \varphi/m_e$, where ${\vec v}$ is the
fluid velocity, $n$ is the fluid particle density, $m_e$ is the electron
mass, and $\varphi$ is the electrostatic potential. The first term on the
right hand side is the acceleration due to a gradient in electron pressure.
The second term is the acceleration due to the electric field of the excess
charges on the surfaces of the electrode. The electric field drives the
electrons/holes toward the surfaces to cancel out the excess charges.

The inertial term in the above equation can be estimated as $|({\vec
v}\cdot{\vec \nabla}){\vec v}|\sim v^2/L\sim 10^6m^2/s^2\times L^{-1}$. Here,
$L$ denotes the characteristic length scale over which we expect a departure
from uniform flow, so that $\nabla\sim L^{-1}$.  Velocity of the flow in the
simulation reaches $v\sim 10^3m/s \ll v_F$, where $v_F \approx
1.4\times10^6m/s$ is the Fermi velocity of bulk gold. The hydrodynamical
pressure can be calculated from the derivative of
the ground state energy with respective to $r_s$, $P/n = - \frac{r_s}{3}\epsilon'(r_s)$, where
$\epsilon(r_s)$ is the energy per particle.\cite{Giuliani-Vignale} Therefore,
$|{\vec \nabla} P|/m_en \sim \frac{1}{5}v_F^2|{\vec \nabla} n|/n$.  Here, we
included only the ground state energy of a noninteracting electron gas,
$\epsilon_0(r_s) = \frac{3}{5}\epsilon_F$.\cite{note:xc} Let $\delta n$
denote the change of the particle density as a result of the current or the
formation of the surface charges. For the present junctions at these biases,
we find that typically $\delta n/n\lesssim 0.01$. Then, the acceleration due
to the pressure gradient is
\begin{eqnarray}
|\vec{a}_{P}| &=& \frac{|\vec\nabla P|}{m_en}
%\\\nonumber&\sim&
\sim\frac{1}{5}\frac{v_F^2}{L}\frac{\delta n}{n}
\nonumber\\&\lesssim&
  3\times10^9 \frac{m^2}{s^2} \times L^{-1} .
\end{eqnarray}

To estimate the magnitude of the electrostatic acceleration, we treat the
layer of charge induced on the facing surfaces of the electrodes (as
illustrated in Fig.\ref{fig:nz_contour}c) as an infinite uniformly charged
wire. The electric field of the wire is given by $|{\vec
\nabla}\varphi|=\lambda/2\pi\varepsilon_0 L$, where $\lambda$ is the linear
density of excess charge, $L$ is the distance to the wire, and
$\varepsilon_0$ is the vacuum permittivity. The linear charge density is
calculated by averaging the charge density difference $e\delta n$ between the
configuration with and without the current flow over the layer in which the
charge accumulated at the contact-electrode interfaces. For the same
junctions, we find $\lambda \sim 0.016\ e/{\rm \AA}$. The acceleration of
charges due to this electric field can be calculated as
\begin{eqnarray}
|\vec{a}_{el}| &=& \frac{e}{m_e}\frac{\lambda}{2\pi\varepsilon_0L}\\  
\nonumber
        &\sim& 5\times10^{11}\frac{m^2}{s^2}\times L^{-1} .
\end{eqnarray}

These crude estimates imply that over the same distance $L$
\begin{equation}
\label{eq:ordering}
|({\vec v}\cdot{\vec \nabla}){\vec v}| < |{\vec a}_P| < |{\vec a}_ 
{el}| ,
\end{equation}
which suggests that the hydrodynamic pressure gradients dominate over the
inertia of the fluid (the flow is subsonic and compressible), while the
maximum electrostatic force due to the surface charges is comparable to or
larger than the pressure gradient force before the surface charge has been
passivated. Therefore, it is plausible that the lateral flow observed in the
simulation is primarily of electrostatic origin. In different junction
geometries or in a different conductance regime, the ordering in
Eq.~\ref{eq:ordering} may be different.

We have also carried out a similar simulation using a parabolically- shaped
constriction that resembles a quantum point contact in the 2D electron gas.
At a similar bias as in the non-parabolic junctions, we find similar surface
charge accumulation along the boundaries of the electrodes in the vicinity of
the contact. We believe the analysis we have provided on the surface charge
formation applies to this case as well. To the best of our knowledge, the
accumulation of the surface charges has not been reported in adiabatic
quantum point contacts before. It would thus be interesting to develop
experimental techniques to explore the surface region of the quantum point
contact in a 2D electron gas and the charge accumulation that we observe in
our simulations.

\section{Conclusions}
\label{sec:conclusions}

In this paper, we have used the microcanonical approach~\cite
{DiVentra04p8025} to study the time-dependent current flow morphology and the
charge distribution in discharging nanojunctions represented by both a
jellium model and pseudopotentials. We have showed that the electron flow in
the nanojunctions exhibits hydrodynamic features analogous to a classical
fluid. We have found that in the atomic case the current flow evolves into
wedge-shaped pattern flanked by stagnant zones. The flow develops nonlaminar
features including eddy currents. We suggest that the ionic lattice at the
junction plays the role of ``obstacles'' in the dynamics of the electron
liquid, with consequent development of these features. We have also
demonstrated that excess surface charges accumulate dynamically along the
electrodes. In addition, we have observed that for a period of time, there is
strong current flow in the transverse direction. By employing an
order-of-magnitude argument we suggest that this flow is driven by both
hydrodynamic forces due to the electron pressure, and electrostatic forces
due to the surface charge distributions. The latter forces dominate the
initial dynamics in the junctions at hand.

The present and a previous study\cite{Bushong05p2569} demonstrate that the
microcanonical approach combined with time-dependent density-functional
theory can be used to probe the transient behavior of the current in
nanojunctions such as atomic-scale point contacts or molecular
junctions.~\cite{Cheng06p155112} The present approach supplements existing
methods that are based on the static scattering picture and provides another
tool to studying relatively unexplored nanoscale transport phenomena from
first principles.

The flow patterns we observe in metallic nanojunctions can be generalized to
a number of other systems, such as molecular junctions, although many details
may vary. In view of the recent advances in microscopic imaging techniques of
coherent current flow in quantum point contacts in a 2D electron
gas,\cite{Topinka00} we hope that new experimental work exploring the
behavior of current flow in atomic contacts and molecular junctions will soon
emerge to supplement our studies.

\begin{acknowledgements}
The work was supported by the Department of Energy grant DE- 
FG02-05ER46204.
\end{acknowledgements}

\appendix
\section{Numerical details}
\label{num}

We performed time-dependent density-functional calculations using the program
{\tt socorro}\cite{socorro} and an in-house program which implements TDDFT
within the jellium model. The gold ions were modeled by norm conserving
Hamann pseudopotentials with $6s$ electrons as valence
electrons.\cite{Hamann89p2980} We used the Perdew-Zunger (1981) LDA
exchange-correlation functional.\cite{Perdew81p5048} The atomic calculation
employed a plane-wave basis set, with an energy cutoff of 204 eV, which
corresponds to grid spacing of 0.2\AA. The energy eigenvalues vary by less
than 1\% by increasing the cutoff by 66\%.  In the jellium case, the
calculations were performed using a real-space basis set where the space is
uniformly discretized and the grid spacing is 0.7 \AA. The eigenvalues vary
less than 3\% by decreasing the grid spacing by 66\%. The time evolution
operator is represented using the Chebyshev method,\cite{Tal- Ezer84p3967}
with a time step of $5 \times 10^{-4}$ fs in the atomic case and $5 \times
10^{-3}$ fs in the jellium case.

%\bibliography{flow}

\begin{thebibliography}{59}
\bibitem{explain} Note that a quasi-steady state may not occur in a  
nanoscale junction, even in the
presence of this fast momentum relaxation mechanism, if nonlaminar  
flow (like the one we observe in this work)
occurs (see also Ref.~\onlinecite{BGD}).
\bibitem[{not({\natexlab{a}})}]{note:vis}
\bibinfo{note}{Note that in practical calculations, the electron  
liquid can
   have a nonvanishing numerical viscosity due to the discrete grid  
spacing.}
\bibitem[{not({\natexlab{b}})}]{note:streamline}
\bibinfo{note}{The nonuniformity of the density of stream lines is  
partially
   due to the fact that we have calculated the streamlines using  
random initial
   points. The lines become denser when two or more initial points  
belong to
   nearby streamlines.}
\end{thebibliography}

\end{document}